\documentclass[npg, ms]{copernicus}

\begin{document}

\nolinenumbers

\title{Relativistic surfatron process for Landau resonant electrons  in radiation belts}

\author[1]{A. Osmane}
\author[2]{A. M. Hamza}

\affil[1]{Aalto University, Radio Science and Engineering, Espoo, Finland}
\affil[2]{University of New Brunswick, Physics Department, Fredericton, New Brunswick, Canada}


\runningtitle{Radiation belts surfatron acceleration}\runningauthor{A.~Osmane
and A.~M.~Hamza}

\received{22 July 2013}
\revised{9 October 2013}
\accepted{12 December 2013}

\correspondence{Adnane Osmane\\ (a.osmane@unb.ca)}

\received{}
\pubdiscuss{} 
\revised{}
\accepted{}
\published{}


\firstpage{1}

\maketitle  

\begin{abstract}
Recent theoretical studies of the nonlinear wave-particle
interactions for relativistic particles have shown that Landau resonant
orbits could be efficiently accelerated along the mean background magnetic
field for propagation angles $\theta$ in close proximity to a critical
propagation $\theta_\textrm{c}$ associated with a Hopf--Hopf bifurcation
condition. In this report, we extend previous studies to reach greater
modeling capacities for the study of electrons in radiation belts by
including longitudinal wave effects and inhomogeneous magnetic fields. We
find that even though both effects can limit the surfatron acceleration of
electrons in radiation belts, gains in energy of the order of 100\,keV,
taking place on the order of ten milliseconds, are sufficiently strong for the
mechanism to be relevant to radiation belt dynamics.
\end{abstract}


\introduction  
The following report aims at extending a theoretical model for wave-particle
coherent interaction \citep{Osmane12, Osmane12b} to characterize electron
dynamics in the Earth's radiation belts. Electrons with energies in the MeV
range have been frequently measured in the inner and outer component of the
radiation belt. An increase in relativistic electron flux observed for short
time scales (from few hours to few days) and in correlation with an increase
of magnetic activity during the recovery phase of geomagnetic substorms
\citep{Friedel02, Obrien03}. Not only relativistic electrons constitute a
threat to satellites and spacecrafts in orbit, but their production has to be
understood in order to account for the magnetospheric energy budget. Thus
far, numerous theoretical models have been proposed. They can be grouped into
two categories: mechanisms relying on radial transport alone as well as those
that rely on internal mechanisms such as wave-particle interactions and
recirculation models (see, e.g.,~\citealp{Friedel02} and reference therein,
\citealp{Albert02, Shprits06, Summers07}). Radial transport is often
described as a diffusion mechanism driven by the fluctuations in the large
scale magnetosphere electric and magnetic fields. As the particles spread
from the outer magnetosphere to smaller equatorial radial distances, $L$, the
first two adiabatic invariants are conserved while the third one is violated,
resulting in an increase of energy \citep{Kulsrud}. On the other hand, local
wave-particle interactions, and other local mechanisms accelerate particles
already present in the inner magnetosphere. Even though it is accepted that
radial diffusion is an important transport mechanism, particle energization
on time scales shorter than the drift period predicted by radial diffusion
has lead to hold wave-particle interaction responsible for a number of
observables. Events occurring on time scale of days are believed to originate
from electromagnetic cyclotron waves and whistler waves through pitch angle
scattering, whereas the more intense and monochromatic chorus waves are
believed to be the source of the strong energy increase occurring on the
shortest time scales \citep{Obara02, Thorne05, Albert02, Summers07}. More
recently, observations of peaks of phase space density, in contradiction with
inward radial diffusion, have shown that wave-particle interactions are
dominant mechanisms \citep{Green04, Horne05, Chen07}.
 
Aside from the outstanding problem of flux enhancement of relativistic
electrons in the radiation belt, wave-particle interaction could also be
proven to hold a decisive role in a number of other magnetospheric problems.
Among them lies the generation of relativistic electron microbursts observed
in association with VLF chorus waves \citep{Lorentzen01, Summers07,
Hikishima10} as well as precipitation rates of electrons entering the loss
cone in pitch angle.

Perhaps more importantly, the most recent waveforms measured in the radiation
belts have revealed additional motivating reasons to consider the
wave-particle interaction as a dominating energy-momentum exchange mechanism
in radiation belt problems. Large-amplitude, monochromatic, obliquely
propagating, and bursty waveforms were not only repeatedly measured in the
radiation belt \citep{Catell08, Kellog10, Kersten11, Wilson11}, but appeared
correlated with electron energization \citep{Wilson11} as well as
relativistic microbursts events \citep{Kersten11}. The correlation between
chorus waves and electron energization in the radiation belts is not recent,
but it is suspected that if such waveforms were more commonly present in the
radiation belts they could be the dominant trigger responsible for the
energization of electrons on short timescales. A study by \citet{Yoon11} has
shown that if one solves the plasma equations self-consistently, such
waveforms were indeed capable of accelerating electrons on kinetic time
scales consistent with the observations. Even though our study lacks the
levels of self-consistency provided by the numerical method developed by
\citet{Yoon11}, we will show hereafter that we arrive at similar conclusions
if we choose parameters consistent with the radiation belt-measured
waveforms.

The large-amplitude wave forms are observed with a longitudinal component and
the analysis above needs to be conducted with the addition of this
compressive electric component. Whereas the addition of the electrostatic
field with the same phase as the electromagnetic components of the fields
would result in the same condition for the surfatron process, a difference in
phase would shift the Hopf bifurcation and have non-trivial effects that need
to be scrutinized. Moreover, radiation belt electrons are confined in the
magnetic field of the Earth, and one must take into account the effect of
field inhomogeneities. We therefore proceed in this in this report by
modifying a previously derived dynamical system \citep{Osmane12} to allow a
study of relativistic electrons in radiation belts.

\section{Longitudinal Effects}
We follow the procedure described in \citet{Osmane12, Osmane12b} for the derivation of a dynamical system to study the interaction of an ion with an obliquely propagating wave composed of a transverse and longitudinal component. As shown in figure 1, the electromagnetic wave is composed of a transverse component along the $(\hat{x}, \hat{y})$ plane and a longitudinal component along the $\hat{z}$ direction superposed onto a background magnetic field $B_0$ in the $(\hat{y}, \hat{z})$ plane :
\begin{equation}
\textbf{E}(\textbf{x},t)= \delta \textbf{E}_{em}(\textbf{x},t)+ \delta \textbf{E}_{k}(\textbf{x},t)\nonumber
\end{equation}
\begin{equation}
\textbf{B}(\textbf{x}.t)=\textbf{B}_0+\delta \textbf{B}(\textbf{x},t)
\end{equation}
We write the longitudinal component as a function of parameters $\eta$ and $\Psi$ as follows : 
\begin{equation}
\delta \textbf{E}_{k}(\textbf{x},t)=\eta \delta E \sin(kz-\omega t+\Psi) 
\end{equation}
The parameter $\eta$ quantify the amplitude of the longitudinal component of the wave with respect to the electromagnetic component. Setting $\eta=1$ would therefore result in having equal electric field amplitudes parallel and perpendicular to the wave vector \textbf{k}. Setting $\eta=0$ recovers the case treated in \citep{Osmane12b}. The parameter $\Psi$ is a phase difference between the longitudinal component and the transverse component. It is added for the sake of completeness.
The dynamical system equation is therefore not fundamentally modified. The difference reside in the addition of an electric field component along the $z$ component of the Lorentz force. It is easy to show that the dynamical system takes the following form \citep{Osmane12b} (see appendix for detailed derivation of the purely transverse case) : 
\begin{equation}
\label{eq:ds_in_ps2}
\left\{ 
\begin{array}{l l} 
\dot{p}_x'=\Omega_0 p_y'\cos(\theta)-\Omega_1p_z' \cos(kz')  +\Omega_0 (p_z'+p_\phi) \sin(\theta)\\
\dot{p}_y'=-\Omega_0p_x'\cos(\theta)+\Omega_1p_z' \sin(kz')\\
\dot{p}_z'= -p_x'\Omega_0 \sin(\theta)+\Omega_1 \frac{n^2-1}{n^2} (p_x\cos(kz') -p_y'\sin(kz')\\
\hspace{6mm}-\frac{1}{n^2}\Omega_1\eta p_z'\sin(kz'+\Psi)+\Omega_1 \frac{n^2-1}{n^2}\eta p_\phi\sin(kz'+\Psi)\\
 \dot{z}'=p_z'v_\Phi/p_\Phi\\
\end{array} 
\right.
\end{equation} 
for the dynamical gyrofrequencies $\Omega_0=\frac{e B_0}{m\gamma c}$ and  $\Omega_1=\frac{e \delta B}{m\gamma c}$, relativistic momentum $p_i=m\gamma v_i'$, refractive index $n=\frac{c}{v_\phi}$, phase-speed $v_\Phi=\omega/k$ and dots indicating time derivatives.  
The time evolution of the dynamical gyrofrequency can then be written as : 
\begin{equation}
\label{eq:omega_0_dot_3}
\frac{d\Omega_0}{dt}=\frac{-\Omega_0\Omega_1 p_\Phi}{m^2\gamma^2c^2}\bigg{(}p_x'\cos(kz')-p_y'\sin(kz')+\eta(p_\phi+p_z')\sin(kz'+\Psi)\bigg{)}
\end{equation} 
We can now proceed by studying the dynamical system properties in terms of the fixed points and their stability as well as the dependence on wave parameters such as wave obliquity $\theta$ and normalized wave amplitude $\delta_1=\delta B/ B_0$. More importantly we would like to know whether the surfatron acceleration mechanism, the processes by which a particle is trapped along the wave vector and accelerated uniformly along the background magnetic field by the parallel electric field \citep{Osmane12b}, is also available when a longitudinal component is added. The surfatron is a trapping effect, and is therefore only achievable if a parallel electric field of sufficiently large amplitude is present. If the propagation of the wave is parallel, the parallel electric field is zero, and neither trapping along the wave vector nor acceleration along the background magnetic field is possible. We therefore assume that the longitudinal component can for various parameters either enhance or destroy the surfatron.\\ 
A quick look at the dynamical system (\ref{eq:ds_in_ps2}) shows that the fixed point for the electromagnetic case ($p_x'=p_z'=0, p_y'=-p_\phi\tan(\theta), kz'=0$) exist for $\Psi=0$. For $\Psi \neq 0$ no fixed points exist. Hence, one would expect the acceleration mechanisms associated with Hopf-Hopf bifurcation to be available for sufficiently large wave amplitude capable of trapping when $\Psi=0$. Whenever $\Psi \neq 0$, one can think of the longitudinal component as a perturbation to uniformly accelerated particles (orbits). Hence, particles (orbits) could still be energized but the longitudinal component could break the locking eventually.  Additionally, if the longitudinal component of the wave, for given parameters $\eta$ and $\Psi$, cancels the parallel component of the electric field with respect to the background magnetic field due to the electromagnetic component, no uniform acceleration should take place. This condition can be written as follows : 
\begin{equation}
 \delta \mathbf{E} \cdot \mathbf{b_0}= -\delta E_y\sin(\theta)+\delta E_k \cos(\theta)=0
\end{equation}
Setting $\Psi=0$, one finds that this condition translates to $\eta=-\tan(\theta)$ for which neither trapping nor uniform acceleration should be possible. We therefore expect the lack of a parallel electric field component to translate into unstable orbits for linear perturbation around the fixed points. In the next section we apply the stability analysis to the dynamical system (\ref{eq:ds_in_ps2}) for the fixed point ($p_x'=p_z'=0, p_y'=-p_\phi\tan(\theta), kz'=0$).

\subsection{Stability Analysis}
We now proceed similarly as for the electromagnetic case \citep{Osmane12b} to quantify the linear stability of the fixed point\footnote{The fixed point are located by setting the four equations of motions as zero and resolving the remaining algebraic equations.}. In order to do so we linearize the dynamical system and assume a perturbation of the form $\sum_{i=1}^{4}\mathbf{\xi_i}e^{\lambda_i t}$, for which an imaginary eigenvalue translates into marginally stable orbits, negative eigenvalues into linearly stable orbits, and positive eigenvalues linearly into unstable orbits. Computing the resulting Jacobian evaluated at the fixed point $\textbf{J}=\frac{\partial F}{\partial x_i}\bigg{|}_{\textbf{x}_0}$ for the longitudinal case we find the following matrix: 
\large{\[J= \left( \begin{array}{cccc}
0 &a&b &0 \\
c & 0 & 0&0 \\
d& 0 & 0&e\\
0&0&f&0 \end{array} \right).\]}\normalsize{
for the parameters $\delta_1$, $\delta_2=\frac{m\omega c}{e B_0}$, $\theta$ and $\eta$ : 
\begin{equation}
a=\frac{\cos(\theta)}{\delta_2\gamma_0}(1-\frac{\tan(\theta)^2}{n^2-1}) 
\end{equation}
\begin{equation}
b=\frac{\mp\delta_1+\sin(\theta)\frac{n^2}{n^2-1}}{\delta_2\gamma_0}
\end{equation}
\begin{equation}
c=- \frac{\cos(\theta)}{\delta_2\gamma_0} 
\end{equation}
\begin{equation}
d=(-\sin(\theta)\pm\delta_1\frac{n^2-1}{n^2})\frac{1}{\delta_2\gamma_0} 
\end{equation}
\begin{equation}
e=\mp\frac{\delta_1}{\delta_2}\frac{n^2-1}{n^2}(\tan(\theta)+\eta)-\lambda
\end{equation}
\begin{equation}
f=\frac{1}{\gamma_0}=\sqrt{1-\frac{v_\Phi^2}{c^2}(1+\tan^2(\theta))}
\end{equation}
Once again the dark $\pm$ and $\mp$ correspond to the fixed points components for $Z=(0, \pi)$, that is the upper sign for $Z=0$ and the lower one for $Z=\pi$. Hence, all four fixed points are represented in this Jacobian matrix and their stability can be analyzed by choosing the right $\pm$ symbols. In order to find the eigenvalues, we need to solve the characteristic polynomial given by the following expression }:

\[ \chi(\lambda) = \left| \begin{array}{cccc}
-\lambda &\frac{\cos(\theta)}{\delta_2\gamma_0}(1-\frac{\tan(\theta)^2}{n^2-1})  & \frac{\mp\delta_1+\sin(\theta)\frac{n^2}{n^2-1}}{\delta_2\gamma_0}  & 0\\
- \frac{\cos(\theta)}{\delta_2\gamma_0}  & -\lambda & 0&0 \\
(-\sin(\theta)\pm\delta_1\frac{n^2-1}{n^2})\frac{1}{\delta_2\gamma_0} & 0 & \mp\frac{\delta_1}{\delta_2}\frac{n^2-1}{n^2}(\tan(\theta)+\eta)\\
0&0&\frac{1}{\gamma_0}&-\lambda \end{array} \right  |.\] 
\normalsize{}
A little algebra results in the following bi-quadratic expression : 
\begin{equation}
\chi(\lambda)=\lambda^4+\zeta_1\lambda^2+\zeta_2=0
\end{equation}
with the values $\zeta_1$ and $\zeta_2$ given by the following expressions : 
\begin{eqnarray}
\zeta_1&=&\frac{\delta_1}{\delta_2\gamma_0}\frac{n^2-1}{n^2}(\tan(\theta)+\eta)\nonumber\\
  &+&\frac{1}{\delta_2^2\gamma_0^2}\bigg{(}\delta_1\frac{n^2-1}{n^2}+1\mp2\delta_1\sin(\theta)\bigg{)}\nonumber\\
\nonumber\\
 \zeta_2&=&\frac{\delta_1}{\delta_2^3\gamma_0^5}\frac{n^2-1}{n^2}(\sin(\theta)\cos(\theta)+\eta\cos(\theta)^2)\nonumber 
\end{eqnarray}
One can compare the Jacobian matrix as well the characteristic equation for the purely electromagnetic case with the expressions above for an additional longitudinal component with $\Psi=0$. It is clear that minimal differences arises as denoted in the appearance of a factor of $\eta$ in the Jacobian and the characteristic equation. The characteristic equation has once again four roots given by the following equation :
\begin{equation}
\lambda_{1,2,3,4}=\pm\sqrt{\frac{-\zeta_1 \pm \sqrt{\zeta_1^2-4\zeta_2}}{2}}
\end{equation} 
Figure (\ref{eig2}) shows the dependence of all four eigenvalue solutions for typical parameters relevant to space plasmas on the propagation angle. It is clear that the Hopf-Hopf bifurcation takes place once again for parameters resulting in $\gamma_0=0$. That is, whenever parameters are such that $n^2-1=\tan^2(\theta)$, the fixed point evolves from marginally stable to linearly unstable when we add a longitudinal component with $\psi=0$. Additionally, a second bifurcation takes place when $\eta=-\tan(\theta)$.  Setting $\eta=-\tan(\theta)$ in the coefficients $\zeta_1$ and $\zeta_2$ results in the following characteristic equation : $\lambda^4+\zeta_1\lambda^2=0$. With $\zeta_1>0$ for $\delta_1\sim O(1)$, it is clear that two eigenvalues are null and two eigenvalues are imaginary. As noted in the previous section, this expression denotes a null parallel electric field resulting in the destructive interference of the parallel longitudinal component and the parallel electromagnetic component. We now investigate the nonlinear effects of the longitudinal components on the surfatron process for various parameters $\eta$ and $\psi$.

\subsection{Landau Resonant Orbits} 
In this section we determine whether the addition of the longitudinal component enhances or prevents the uniform acceleration for orbits caught in the basin of attraction centered at Landau resonance.
As noted in the stability analysis, a Hopf-Hopf bifurcation does indeed take place when a longitudinal component is added. The main difference in the linear stability around the fixed points resides in the addition of a parallel electric field capable of canceling the electromagnetic component parallel to the background field. Hence, whenever $\eta=-\tan(\theta)$, the parallel component of the electric field is zero and the surfatron process can not take place. Indeed, choosing the parameter $\eta$ to coincide with $\sqrt{n^2-1}$ results in destroying the Hopf-Hopf bifurcation to a single Hopf bifurcation (one pair of imaginary eigenvalues crossing the real plane instead of two pairs). For such parameter the surfatron process is expected to not be applicable because a parallel electric field causing the uniform acceleration is now set to a null value.\\
Figure (\ref{electro_trapped}) shows two orbits for $\eta > -\tan(\theta)$ (on the right panel) and $\eta < -\tan(\theta)$ (on the left panel). A transition from  untrapped to trapped orbit is observed as we evolve the parameter $\eta$. For $\eta=\tan(\theta)$ the particle located sufficiently close to the fixed point (Landau resonant velocity) is trapped, but small perturbation results in untrapped orbits. \\
In figures (\ref{electro_surf}), the four panels represent a seeded particle with energy of the order of $100$ keV but for $\theta=40^o$(up and left), $\theta=55^o$(up and right), $\theta=70^o$(down and left) and $\theta=85^o$(down and right). The particle gains a maximum amount of energy for $\theta=70^o$, which corresponds to a propagation angle close to $\theta_c$ and to the surfatron process. The longitudinal component for parameters  $\delta_1=0.045, v_\phi=0.33 c, \eta=-1, \delta_2=0.1$ enhances the parallel electric field component and results in smaller range of values in $\delta_1=\delta B/ B_0$ for which the surfatron process is accessible. In this case the ratio $\delta_1$ is of the order of  $4\%$. This order of magnitude for the electromagnetic wave component is comparable to large-amplitude bursty waves observed in the radiation belts  \citep{Catell08, Kellog10, Kersten11, Wilson11}.\\
However, before concluding that the longitudinal component preserves the surfatron mechanism for $\theta=\theta_c$ we need to evaluate the effects of the phase difference $\Psi$. In the figures (\ref{electro_surf}) the parameter $\Psi$ has been set to zero. Yet, when $\Psi \neq 0$, the dynamical system possesses no fixed points and the acceleration observed for the surfatron process should not arise uniformly. Panels in figure (\ref{effectpsy}) show particles for $\eta=1, \delta_1=0.06, \delta_2=0.1, n^2=9, \theta=-71^o$ and three different values for $\Psi = -\pi/4, 0$ and $\pi/4$. It is seen that depending on the phase difference, the surfatron process can still take take place for sufficiently long time to energize the particle. As for the purely electromagnetic case, a charged particle in this field would gain a significant amount of energy (from keV levels to MeV) on small kinetic time scales $\omega t \sim 0.1\Omega_0 t \sim 10$. Hence a particle can be energized in such a field on time scales of the order of the $1/100$ of a second for a wave frequency $\omega\sim 3$kHz. Since the acceleration takes place on very small time scales, inhomogeneous effects should not prevent the mechanism entirely.  We therefore conclude this section by suggesting that the large-amplitude electromagnetic waves observed in the radiation belts can energize particles efficiently on kinetic time scales for propagation angles close to the critical Hopf-Hopf bifurcation value $\theta_c$. If the propagation angle is not sufficiently close to $\theta_c$, then the particle will just oscillate back and forth in the potential of the electric field and no significant gain in energy should be observed. We now proceed in the next section by quantifying the inhomogeneous effects on the surfatron herein described.

\section{Inhomogeneous Magnetic Field Effects on the Surfatron}
In this section we want to include the effects of a non-homogeneous magnetic field on the acceleration process described in the previous section. We first discuss the motion of a particle in an inhomogeneous magnetic field with no wave-particle interaction. We then provide an approximation for the time scales for which inhomogeneous effects can result in surfatron breaking and a discussion on numerical integration of a particle trajectory interacting with a large-amplitude electromagnetic wave in an inhomogeneous magnetic field. 
\subsection{Particle orbits in inhomogeneous magnetic field}
Relativistic electrons trapped in the radiation belts bounce back and forth along the  (approximately) dipolar magnetic field of the Earth. Before addressing the more complicated motion of relativistic electrons bouncing back and forth in the Earth's magnetic field and at the same time interacting with an obliquely propagating wave, we would like to quantify the impact of the magnetic field inhomogeneities on the relativistic motion. Assuming the magnetic moment :
\begin{equation}
\mu=\frac{p_\perp^2}{2mB}=\frac{mc^2(\gamma^2-1)\sin(\alpha)^2}{2B}
\end{equation}
is an adiabatic invariant, we can derive the forces due to the magnetic field inhomogeneities as follows.\\
The magnetic force perpendicular to the magnetic field can be deduced from the conservation of the magnetic moment : 
\begin{equation}
\dot{p}_\perp=m\mu \dot{B}/p_\perp.
\end{equation}
Replacing $\mu$ in terms of $B$ and $p_\perp$, and $\dot{B}=v_\parallel \nabla_\parallel B$, we find the following expression for the force perpendicular to the magnetic field : 
\begin{equation}
\label{p_perp_inhomo}
\dot{p}_\perp=\frac{p_\parallel p_\perp \nabla_\parallel B}{2mB\gamma}
\end{equation}
Using the above equation and assuming that the energy of the particle is conserved to first order in $\mu$ for a particle moving in an inhomogeneous magnetic field, we can write : 
\begin{equation}
\gamma \dot{\gamma}=p_\parallel \dot{p}_\parallel+ p_\perp\dot{p}_\perp=0
\end{equation}
Therefore, the equation of motion in the parallel direction can be written as : 
\begin{eqnarray}
\label{p_para_inhomo}
\dot{p}_\parallel&=&- p_\perp\dot{p}_\perp/p_\parallel \nonumber\\
          &=&-\frac{p_\perp^2 \nabla_\parallel B}{2m B\gamma} \nonumber \\
          &=&-\frac{\mu}{\gamma} \nabla_\parallel B
\end{eqnarray}
Setting $\gamma=1$ in equations (\ref{p_para_inhomo}) and (\ref{p_perp_inhomo}) recovers the expressions for non-relativistic particles \citep{Bell81}. We can show that similarly to the non-relativistic case, the conservation of magnetic moment results in magnetic trapping. Assuming a slab geometry for the magnetic field as shown in (\ref{topology3}) for $z\sim s_\parallel$, i.e. the component along the parallel coordinate, and the magnetic field of the form $B(s_\parallel)=B_0(1+\frac{9}{2}s_\parallel^2/R^2)$ to mimic the dipolar field, for which $R$ is equal to the Earth's radius, we can solve  both equations (\ref{p_para_inhomo}) and (\ref{p_perp_inhomo}). Hence, replacing the expression for $B(s_\parallel)$ in  (\ref{p_para_inhomo}), results in the following equation : 
\begin{equation}
\frac{d^2 s_\parallel}{dt^2}+\frac{9\mu B_0}{m\gamma^2 R^2} s_\parallel=0.
\end{equation}
The solution of the above equation is therefore of the form $s_\parallel \sim \cos(\sqrt{\frac{9\mu B_0}{m\gamma^2 R^2}}t)$. Replacing the expression for the magnetic field in equation (\ref{p_perp_inhomo}) for the perpendicular momentum, we find the following differential : 
\begin{equation}
\frac{dp_\perp}{p_\perp}=\frac{9}{2R^2}d(s_\parallel^2)
\end{equation}
with a perpendicular momentum solution 
\begin{equation}
p_\perp=p_{\perp 0}e^{(\frac{9}{2R^2}s_\parallel^2)}. 
\end{equation} 
Replacing the solution for $s_\parallel$ in the above equation provides for a complete solution for the particle motion in the inhomogeneous field $B(s_\parallel)=B_0(1+\frac{9}{2}s_\parallel^2/R^2)$. We see that the particle oscillates back and forth  along the parallel direction, while the perpendicular momentum increases as the particle reaches regions of larger magnetic field strength corresponding to $s_\parallel \sim R$. 

\subsection{Surfatron breaking due to inhomogeneous B field effect.} 
Numerous effects can cause the breaking of the surfatron process : dispersive wave-effects, dissipation of the wave amplitude, inhomogeneous magnetic fields damping the acceleration along the field line, or simply the result of precipitation into the atmosphere. Because of the slow time-scales upon which the surfatron process becomes interesting to sustain particle precipitation, and since other effects would take place on longer time-scales,we now focus solely on the inhomogeneous effect. That is, we want to obtain time scales for which the surfatron would not be prevented by field inhomogeneities. \\
The surfatron results in the parallel acceleration of a particle caused by the parallel component of the electric field. As demonstrated in the previous section, magnetic field inhomogeneities result in a $-\mu \nabla B$ force that can reduce the surfatron process, in the same way that the parallel electric field from a longitudinal component can prevent parallel acceleration. Hence, when $e\delta E_\parallel \sim \frac{\mu}{\gamma} \nabla B$ parallel acceleration becomes marginal. This condition translates as follows : 
\begin{equation}
\frac{s_\parallel}{R} \sim \frac{\delta_1}{n} \frac{\Omega_0 \gamma R}{(\gamma_0^2-1)\sin(\alpha_0)^2 c} 
\end{equation}
for the parameters $\delta_1= \delta B/ B_0$ and $n=c/v_\phi$, the gyrofrequency $\Omega_0=eB_0/mc$ as well as the pitch-angle $\alpha_0$ and Lorentz factor $\gamma_0$ of the particle at the equatorial region. Setting $n=3$, $\delta_1 \sim 0.01$, $R \sim 6000$ km, $\Omega_0 \sim 3\times 10^4$ for a particle with an initial energy of the order of $100$ keV and pitch-angle of $45^o$, we find $s_\parallel \sim 10^4$ km. Therefore, we can conclude that a particle would gain energy of the order of $W \sim e\delta E_\parallel s_\parallel$, which corresponds to a gain in energy of the order of 100 keV for an electron interacting with an electric field of $100$ mV/m. This approximation is comparable to energization of electrons reported by \citep{Artemyev13} who found gains of the order of $80-100$ keV for particles going through several Landau resonance in an inhomogeneous field. The difference with our result hereafter, however, is that the particle gains energy during the time of one Landau resonance, making the process much more efficient, even though less probable.\\

We now write a dynamical system for a relativistic charged particle interacting with an obliquely propagating wave in an inhomogeneous field. In order to make the set of equations more transparent to the reader we write them in a coordinate axis for which $\hat{z} \parallel \hat{b}$, that is the background magnetic field is parallel to the $z$-axis. We denote $\hat{y}=\hat{\perp}_1$ and $\hat{x}=\hat{\perp}_2$. Rewriting the magnetic field in terms of this coordinate system we obtain :

\begin{equation}
\left\{ 
\begin{array}{l l} 
\delta B_x =  \delta B\sin(\Phi) \\ 
\delta B_y =  \delta B\cos(\Phi)\cos(\theta) \\ 
\delta B_\parallel =  -\delta B\cos(\Phi)\sin(\theta) \\ 
\end{array} \right.
\end{equation}

and similarly, using Faraday's laws, we obtain the following components for the electric field : 

\begin{equation}
\left\{ 
\begin{array}{l l} 
\delta E_x =  -v_\Phi\delta B\cos(\Phi)/c \\ 
\delta E_y =  v_\Phi\delta B\sin(\Phi)\cos(\theta)/c \\ 
\delta E_\parallel =  -v_\Phi\delta B\sin(\Phi)\sin(\theta)/c \\ 
\end{array} \right.
\end{equation}

for the phase $\Phi=k_\parallel z+k_{\perp1}y-\omega t=kz-ky-\omega t$. We choose the background magnetic field to be written as $\mathbf{B_0}=-yB_0g'(z)\hat{y}+B_0g(z)\hat{z}$, for the function $g(z)=1+z^2/R^2$ and its partial derivative with respect to $z$, $g'(z)$, denoting the background magnetic field variation as the particle propagates toward along the field line and away from the equatorial region $z\sim 0$. We can then write the dynamical system in terms of the variables ($p_x, p_y, p_z, z, y, \Omega_0=eB_0/m\gamma c, \Omega_1=e\delta B/m\gamma c$) and the functions $g(z)$ and $q(y,z)=yg'(z)$:
\begin{equation}
\left\{ 
\begin{array}{l l} 
\dot{p}_x=-p_\phi\Omega_1\cos(\Phi)+p_y\Omega_0g(z)-p_z\Omega_0q(y,z)-p_y\Omega_1\cos(\Phi)\sin(\theta)\\ -p_z\Omega_1\cos(\Phi)\cos(\theta)\\
\dot{p}_y=p_\phi\Omega_1\sin(\Phi)\cos(\theta)-p_x\Omega_0g(z)+p_z\Omega_1\sin(\Phi)+p_x\Omega_1\cos(\Phi)\sin(\theta)\\
\dot{p}_z=p_x\Omega_0q(y,z)-p_\phi\Omega_1\sin(\Phi)\sin(\theta)+p_x\Omega_1\cos(\Phi)\cos(\theta)-p_y\Omega_1\sin(\Phi)\\
\dot{y}=p_y v_\Phi/p_\Phi \\
\dot{z}=p_z v_\Phi/p_\Phi
\end{array} 
\right.
\end{equation} 
and 
\begin{eqnarray}
\label{eq:omega_dot_4}
\dot{\Omega}_0 & = & \frac{d}{dt}\bigg{(}\frac{eB_0}{mc\gamma}\bigg{)} \nonumber \\
                            & = & -\bigg{(}\frac{eB_0}{mc\gamma}\bigg{)} \frac{1}{\gamma}\frac{d\gamma}{dt}\nonumber \\
                            &=&-\Omega_0 \frac{p\dot{p}}{m^2\gamma^2c^2}\nonumber \\
                            &=&-\Omega_0 \frac{pc^2}{m^2c^4+p^2c^2}\dot{p}\nonumber\\
                            &=&\frac{-\Omega_0\Omega_1 p_\Phi}{m^2\gamma^2c^2}\bigg{(}\sin(\Phi)(p_y\cos(\theta)-p_z\sin(\theta))-p_x\cos(\Phi)\bigg{)}
                            \end{eqnarray}
We then proceed by normalizing the variables as follows : $p_i/mv_\phi = P_i$, $kz =  Z$, $ky =  Y$, $\Omega_0/\omega =  \delta_3$ and $\omega t = \tau$, and write the dynamical system in terms of the normalized variables $P_x, P_y, P_z, Y, Z, \delta_3$ and the parameters $\delta_1, \delta_2$ and $n$ as previously defined. 

\begin{equation}
\left\{ 
\begin{array}{l l} 
\dot{P}_x=-\delta_1\cos(\Phi)/\delta_2+P_y\delta_3g(Z)-P_z\delta_3q(Y,Z)-P_y\delta_1\delta_3\cos(\Phi)\sin(\theta)\\
-P_z\delta_1\delta_3\cos(\Phi)\cos(\theta)\\
\dot{P}_y=\delta_1\sin(\Phi)\cos(\theta)/\delta_2-P_x\delta_3g(Z)+P_z\delta_1\delta_3\sin(\Phi)+P_x\delta_1\delta_3\cos(\Phi)\sin(\theta)\\
\dot{P}_z=P_x\delta_3q(Y,Z)-\delta_1\sin(\Phi)\sin(\theta)/\delta_2+P_x\delta_1\delta_3\cos(\Phi)\cos(\theta)-P_y\delta_1\delta_3\sin(\Phi)\\
\dot{Y}=\delta_2\delta_3P_y\\
\dot{Z}=\delta_2\delta_3P_z\\
\dot{\delta_3}= -\frac{\delta_1\delta_2\delta_3^3 }{n^2}\bigg{(}\sin(\Phi)(P_y\cos(\theta)-P_z\sin(\theta))-P_x\cos(\Phi)\bigg{)}

\end{array} 
\right.
\end{equation}
We can now integrate this dynamical system for parameters relevant to
radiation belt electrons with large-amplitude obliquely propagating waves to
study the surfatron process in an inhomogeneous magnetic field mimicking the
Earth's magnetic field. Figure~\ref{gradB_gamma} show orbits for parameters
$\delta_{1}$\,=\,0.01, $v_{\phi}$\,=\,0.33\,$c$, $\delta_{2}$\,=\,0.1 and
$\theta$\,=\,71.9$^\circ$. The left panel shows the perpendicular component
of momentum $P_{y}$ against the perpendicular position $Y$. The right panel
shows the parallel component of momentum $P_z$ against the parallel position
$Z$. Initially the parallel component of momentum increases because of
surfatron, until the $-\mu\,\nabla\,B$ force becomes \mbox{sufficiently} strong to
break the trapping due to the parallel electric field. The particle gains
energy of the order of 76\,keV during the process. This gain in energy is of
similar order ($\sim$\,100\,keV) as the one computed above for a simple
balance of forces. Even though this gain appears modest, it should be kept in
mind that it takes place on a time scale of the order of $\tau$\,$\sim$\,12,
hence $t$\,$\sim$\,10\,ms for a wave frequency $\omega$\,$\sim$\,3\,kHz. As
the surfatron acceleration is lost, the conservation of the adiabatic
invariant leads to a transfer of energy from the parallel direction to the
perpendicular as noted by the continuous increase of the perpendicular
momentum for $\tau$\,$>$\,12. Since the gain in energy is irreversible, the
particle uplifted by tens of keV can only oscillate back and forth in the
potential of the wave and transfer energy along the perpendicular or parallel
direction to preserve adiabatic invariance.

This point is clearly demonstrated in the right panel of
Fig.~\ref{gradB_gamma}. As the particle is uniformly accelerated, $\gamma$
increases. Once the surfatron is broken, the particle resides in a state of
higher energy. The left panel of the figure shows the three-dimensional orbit
in velocity space $V\,x$, $V\,y$, $V\,z$. For $\tau$\,$<$\,12, the particle
is uniformly accelerated through the surfatron process along the parallel
direction. Once the surfatron is made inoperable, the magnetic field gradient
dictates the particle orbits and conservation of $\mu$ leads to transfer of
energy to the perpendicular direction. As denoted by the left panel of
Fig.~\ref{gradB_gamma}, the gyroradius of the particle increases
($V_{y}$\,$\gg$\,$V_{z}$) as the particle comes out of the surfatron.

Even though the surfatron accelerates particles parallel to the magnetic
field, the inhomogeneous field results in redistributing the energy
perpendicularly to the magnetic field. Hence, such a process, if
statistically common in the radiation belts, could provide for an explanation
to anisotropic distribution resulting in whistler wave turbulence without the
need to resort to cyclotron resonance. Indeed, both resonant and nonresonant
electron whistler instabilities require an initial anisotropy with $K_{\perp}$,
the perpendicular kinetic energy density, to exceed $K_{\parallel}$, the
parallel kinetic energy density by a certain amount. As the waves are being
triggered by the instability and $K_{\parallel}/K_{\perp}$ reaching marginal
stability levels, particles in the tail can be accelerated through the
surfatron, travel toward a region of larger magnetic field and gain greater
gyroradius, bounce back to equatorial region and contribute to the breaking
of the marginal stability state of the whistler turbulence. This back and
forth mechanism could then be stopped by precipitating the particles in the
atmosphere, instead of having them bounce back toward the equatorial region.

\conclusions  
We extended a previous theoretical study \citep{Osmane12b} of nonlinear
wave-particle interactions for the study of electrons in radiation belts by
including longitudinal wave effects and inhomogeneous magnetic fields. We
found that, similarly than for the electromagnetic case in a uniform
background field, the acceleration of particles along the background magnetic
field, for propagation angles in close proximity to a critical propagation
$\theta_c$ and associated with a Hopf--Hopf bifurcation condition, can arise
on sufficiently small timescales to be of relevance to radiation belt
dynamics. Even though longitudinal wave components and inhomogeneous magnetic
fields can limit the surfatron acceleration of electrons in radiation belts,
gains in energy of the order of 100\,keV, taking place on timescales of the order of 10\,ms, 
are sufficiently strong for the mechanism to sustain efficient particle
energization. Future studies will investigate the effect of wave obliquity and field inhomogeneities on
electron distribution functions for parameters consistent with radiation belt
dynamics.

\appendix
\section{Dynamical system derivation for a transverse electromagnetic wave}

Our starting point is the relativistic Lorentz equation for the motion of a particle in an electromagnetic field. The force is therefore written as

\begin{equation}
\frac{d\textbf{p}}{dt}= e\bigg{[}\textbf{E}(\textbf{x},t)+\frac{\textbf{v}}{c} \times \textbf{B}(\textbf{x},t)\bigg{]}
\end{equation}
for a particle of momentum $\textbf{p}=m\gamma\textbf{v}$, rest mass $m$ and charge $e$. The Lorentz contraction factor $\gamma$ is defined as follows : 
\begin{equation}
\gamma=\frac{1}{\sqrt{1-\frac{v^2}{c^2}}}
\end{equation}
In order to avoid dealing with both the velocity and the momentum in the dynamical system, we simply write the equations in terms of the more physically relevant quantity between the two, that is the relativistic momentum $\textbf{p}$ : 
\begin{equation}
\frac{d\textbf{p}}{dt}= e\bigg{[}\textbf{E}(\textbf{x},t)+\frac{\textbf{p}}{m\gamma c} \times \textbf{B}(\textbf{x},t)\bigg{]}
\end{equation}
Similarly, the Lorentz factor can be written as follows : 
\begin{equation}
\label{eq:constraint}
\gamma=\frac{\sqrt{m^2c^4+p^2c^2}}{mc^2}
\end{equation}
The electromagnetic field is superposed onto a background magnetic field $B_0$.
\begin{equation}
\textbf{E}(\textbf{x},t)=\delta \textbf{E}(\textbf{x},t)
\end{equation}
\begin{equation}
\textbf{B}(\textbf{x}.t)=\textbf{B}_0+\delta \textbf{B}(\textbf{x},t)
\end{equation}
The electromagnetic wave vector points in the $\hat{z}$ direction and the background magnetic field lies in the $y-z$ plane. 
\begin{equation}
\textbf{k} \cdot \textbf{B}_0=kB_0\cos(\theta)
\end{equation}
\begin{equation}
\left\{ 
\begin{array}{l l} 
\delta\textbf{E} =  \delta E_x \hat{\textbf{x}} + \delta E_y \hat{\textbf{y}}\\ 
\delta\textbf{B} =  \delta B_x \hat{\textbf{x}} + \delta B_y \hat{\textbf{y}}\\   
\end{array} \right.
\end{equation}
where 
\begin{equation}
\label{deltaBfield}
\left\{ 
\begin{array}{l l} 
\delta B_x= \delta B \sin(kz-\omega t)\\ 
\delta B_y=\delta B \cos(kz-\omega t)\\   
\end{array} \right.
\end{equation}
Faraday's law, expressed in terms of the Fourier components gives the relation between the components of the electromagnetic fields.
\begin{equation}
\label{deltaEfield}
c\textbf{k} \times \delta \textbf{E} (\textbf{k},\omega) =\omega \delta \textbf{B}(\textbf{k},\omega)
\end{equation}
The electric force is therefore written as 
\begin{equation}
\left\{ 
\begin{array}{l l} 
F_{Ex}=e v_\phi \delta B \cos(kz-\omega t)/c\\ 

F_{Ey}=- e v_\phi \delta B \sin(kz-\omega t)/c\\ 

F_{Ez}=0\\
\end{array} \right. 
\end{equation}
for which $v_\phi=\omega/k$ is the phase velocity. Taking the cross product of the momentum and the magnetic field, the magnetic force is written as 
\begin{equation}
\left\{ 
\begin{array}{l l} 
F_{Bx}= \frac{1}{m\gamma c}(p_y eB_0 \cos(\theta)+p_z eB_0 \sin(\theta)-p_ze\delta B_y) \\ 
F_{By}=\frac{1}{m\gamma c}( -p_x eB_0 \cos(\theta)+p_ze\delta B_x)\\   
F_{Bz}=\frac{1}{m\gamma c}( -p_x eB_0 \sin(\theta)+p_xe\delta B_y-p_ye\delta B_x)\\
\end{array} 
\right. 
\end{equation}
We can write the dynamical system equations in terms of the following variables : $p_\Phi=m\gamma v_\Phi$ ; $\Omega_1=e\delta B/mc\gamma$ ; $\Omega_0=e B_0/mc\gamma$ , which results in the following equations : 
\begin{equation}
\left\{ 
\begin{array}{l l} 
\dot{p}_x=p_y\Omega_0\cos(\theta)+(p_\Phi-p_z)\Omega_1\cos(kz-\omega t)  +p_z\Omega_0\sin(\theta)\\
\dot{p}_y=-p_x\Omega_0\cos(\theta)+(p_z-p_\Phi)\Omega_1\sin(kz-\omega t)\\
\dot{p}_z=-p_x\Omega_0 \sin(\theta)+p_x\Omega_1\cos(kz-\omega t) -p_y\Omega_1\sin(kz-\omega t)\\
\dot{z}=p_z v_\Phi/p_\Phi
\end{array} 
\right.
\end{equation} 
In the classical case we have $4$ equations to integrate, the three components of the velocity plus the position coordinate along \textbf{k}. In the relativistic case, for a non zero propagation angle, the energy of the particle is not a constant of the motion, that is, $\dot{\gamma} \neq 0$. Hence, we can think of the relativistic dynamical system as composed of 4 equations, the three components of the momentum plus the position coordinate along \textbf{k}, and one constraint relating $\gamma$ and the momentum components. Without any loss of generality we take the constraint into consideration by writing an equation for the dynamical gyrofrequency :
\begin{eqnarray}
\label{eq:omega_dot}
\dot{\Omega}_0 & = & \frac{d}{dt}\bigg{(}\frac{eB_0}{mc\gamma}\bigg{)} \nonumber \\
                            & = & -\bigg{(}\frac{eB_0}{mc\gamma}\bigg{)} \frac{1}{\gamma}\frac{d\gamma}{dt}\nonumber \\
                            &=&-\Omega_0 \frac{p\dot{p}}{m^2\gamma^2c^2}\nonumber \\
                            &=&-\Omega_0 \frac{pc^2}{m^2c^4+p^2c^2}\dot{p}\nonumber\\
                            &=&-\frac{\Omega_0\Omega_1p_\Phi}{m^2\gamma^2c^2}\bigg{(}p_x\cos(kz-\omega t)-p_y\sin(kz-\omega t)\bigg{)}
\end{eqnarray}
If we define the constant $\delta=\Omega_1/\Omega_0$, it is straightforward to see that 
\begin{equation}
\dot{\Omega}_1=\delta \dot{\Omega}_0
\end{equation}
Since $p_\Phi=p_\Phi(\gamma)$, the time evolution of this quantity is written as :
\begin{eqnarray}
\label{eq:p_phi}
\dot{p}_\Phi&=&mv_\Phi\frac{d\gamma}{dt}\nonumber \\
                     &=&mv_\Phi\frac{p}{\gamma m^2c^2}\dot{p}\nonumber\\
                     &=&-mv_\Phi\gamma\frac{\dot{\Omega}_0}{\Omega_0}
\end{eqnarray} 
We can now eliminate the explicit time dependence of the equations by making a transformation of variables. Even though this transformation corresponds to a translation in the wave frame for low phase-speed of the wave ($v_\phi \ll c$), it does not correspond to a physical frame of reference for phase-speeds similar to the speed of light $v_\phi \sim c$  \footnote{The transformation in the position coordinates along $z$ and $v_z$ are Lorentz transformations in the wave frame, but because we do not also transform the time component into the wave frame ($t'=\gamma (t+v_\phi z/c^2)$, the dynamical system does not correspond to a particle orbit in the wave frame for relativistic regimes.}. The explicit time dependence can therefore be eliminated by the following change of variables : 
\begin{equation}
\label{eq:transformation}
p_x'=p_x, \hspace{2mm}p_y'=p_y, \hspace{2mm} p_z'=\gamma_w(p_z-p_\phi), \hspace{2mm} z'=\gamma_w(z-v_\phi t)
\end{equation}
for the Lorentz factor :
\begin{equation}
\gamma_w=\frac{1}{\sqrt{1-\frac{v_\Phi^2}{c^2}}}
\end{equation}
We can then write the equations of motion in terms of the new variables as follows ;
\begin{equation}
\left\{ 
\begin{array}{l l} 
\dot{p}_x'=\Omega_0 p_y'\cos(\theta)-\Omega_1p_z' \cos(kz'/\gamma_w)/\gamma_w  +\Omega_0 (p_z'/\gamma_w+p_\phi) \sin(\theta)\\
\dot{p}_y'=-\Omega_0p_x'\cos(\theta)+\Omega_1p_z' \sin(kz'/\gamma_w)/\gamma_w\\
\dot{p}_z'/\gamma_w=-\Omega_0p_x'\sin(\theta)+\Omega_1p_x'\cos(kz'/\gamma_w) -\Omega_1p_y'\sin(kz'/\gamma_w)-\dot{p}_\Phi\\
\dot{z}'=p_z'v_\Phi/p_\Phi\\
\end{array} 
\right.
\end{equation} 
If we absorb the Lorentz factor $\gamma_w$ into $p_z'$ and $k$, that is, we write $p_z' \to p_z'/\gamma_w$ and $k \to k/\gamma_w$, and write $\dot{p}_\Phi$ in terms of $(p_x',p_y',p_z',z',\Omega_0)$, we can write the dynamical system as follows : 
\begin{equation}
\label{eq:ds_in_ps}
\left\{ 
\begin{array}{l l} 
\dot{p}_x'=\Omega_0 p_y'\cos(\theta)-\Omega_1p_z' \cos(kz')  +\Omega_0 (p_z'+p_\phi) \sin(\theta)\\
\dot{p}_y'=-\Omega_0p_x'\cos(\theta)+\Omega_1p_z' \sin(kz')\\
\dot{p}_z'=-\Omega_0p_x'\sin(\theta)+\Omega_1(\frac{n^2-1}{n^2})(p_x'\cos(kz') -p_y'\sin(kz'))\\
\dot{z}'=p_z'v_\Phi/p_\Phi\\
\end{array} 
\right.
\end{equation} 
with the refractive index $n^2=c^2/v_\Phi^2$. The magnitude of the momentum is now written as $p'=\sqrt{p_x^{'2}+p_y^{'2}+(p_z^{'}/\gamma_w)^2}$, hence the Lorentz contraction factor also transforms from $\gamma(p) \to \gamma(p')$. 

\begin{acknowledgements}
This work was supported by the Natural Sciences and Engineering Research
Council of Canada (NSERC). One of the authors, A.~M.~Hamza, wishes to
acknowledge CSA (Canadian Space Agency) support. Computational facilities are
provided by ACEnet, the regional high-performance computing consortium for
universities in Atlantic Canada.\hack{\newline} \hack{\newline} Edited by:
T.~Hada \hack{\newline} Reviewed by: P.~Yoon and G.~S.~Lakhina
\end{acknowledgements}

%









%
%


%
%
\begin{figure}[ht]   \centering
   \includegraphics[width=0.45\textwidth]{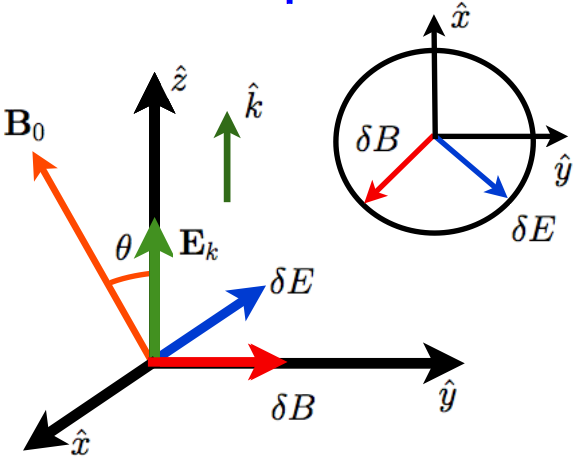}
   \caption{Electromagnetic field configuration. A circularly polarized wave propagating obliquely to a background magnetic field $\mathbf{B_0}$ with a longitudinal component.}
   \label{topology2}
\end{figure}

\begin{figure}[htb] 
   \centering
   \includegraphics[width=0.70\textwidth]{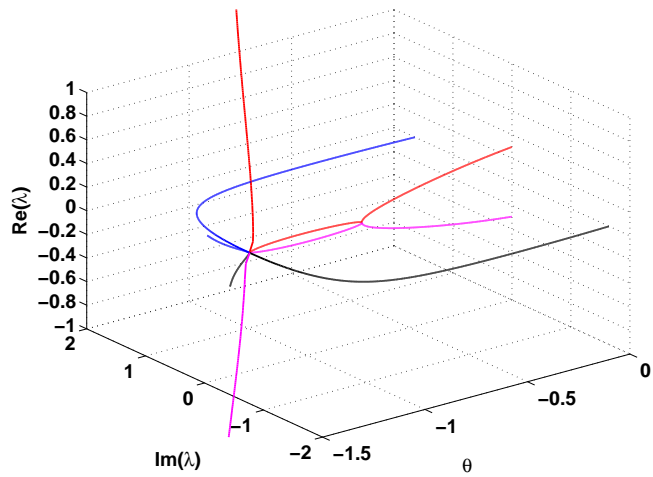}
   \caption{Eigenvalues' dependence on the propagation angle $\theta$ for fixed parameters $\delta_1=0.07$, $\delta_2=0.21$, $n^2=2$, $\eta=0.9$ and the fixed point of component $Z_0=0$. A Hopf-Hopf bifurcation takes place for $\tan(\theta_c)^2=n^2-1$ and a second bifurcation takes place for $\eta=-\tan(\theta)$, corresponding to a null parallel electric field.  The fixed point is stable for $\theta < \theta_c$ and $\eta>-\tan(\theta)$ and unstable for $\theta>\theta_c$ and  $\eta>-\tan(\theta)$.}
   \label{eig2}
\end{figure}

\begin{figure}[ht]
\begin{minipage}[b]{0.40\linewidth}
\centering
\includegraphics[width=\textwidth]{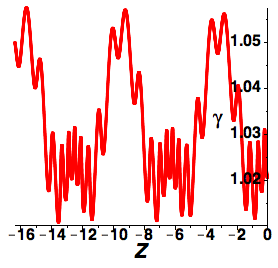}
\end{minipage}
\hspace{0.25cm}
\begin{minipage}[b]{0.40\linewidth}
\centering
\includegraphics[width=\textwidth]{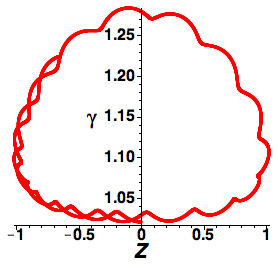}
\end{minipage}
\caption{Orbit for parameters $\delta_1=0.1, v_\phi=0.33 c, \theta=-31^o, \delta_2=.1$ for $\eta > -\tan(\theta)$ (right panel) and $\eta < -\tan(\theta)$ (left panel). A transition from  untrapped to trapped orbit is observed as we evolve the parameter $\eta$.}
\label{electro_trapped}
\end{figure}

\begin{figure}[ht]
\begin{minipage}[b]{0.30\linewidth}
\centering
\includegraphics[width=\textwidth]{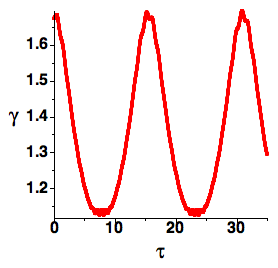}
\includegraphics[width=\textwidth]{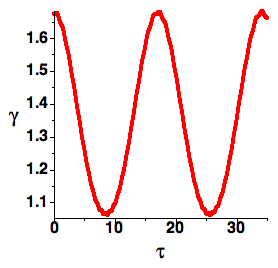}
\end{minipage}
\hspace{0.25cm}
\begin{minipage}[b]{0.30\linewidth}
\centering
\includegraphics[width=\textwidth]{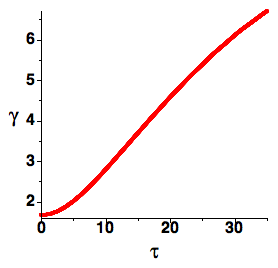}
\includegraphics[width=\textwidth]{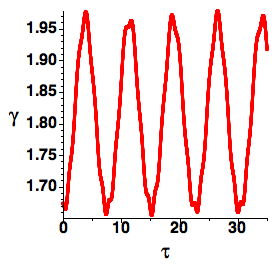}
\end{minipage}
\caption{Orbit for parameters $\delta_1=0.045, v_\phi=0.33 c, \eta=-1, \delta_2=.1$ and  $\theta=40^o$(up and left), $\theta=55^o$(up and right), $\theta=70^o$(down and left) and $\theta=85^o$(down and right). Each case is seeded with a particle of energy of the order of $100$ keV. The particle gains a maximum amount of energy for $\theta=70^o$, which correspond to a propagation angle close to $\theta_c$. The longitudinal  component can enhance the parallel electric field component and result smaller range of values $\delta_1=\delta B/ B_0$ for which the surfatron process is accessible. In this case the ratio $\delta_1 \sim  4\%$.}
\label{electro_surf}
\end{figure}

\begin{figure}[htb] 
   \centering
   \includegraphics[width=1.0\textwidth]{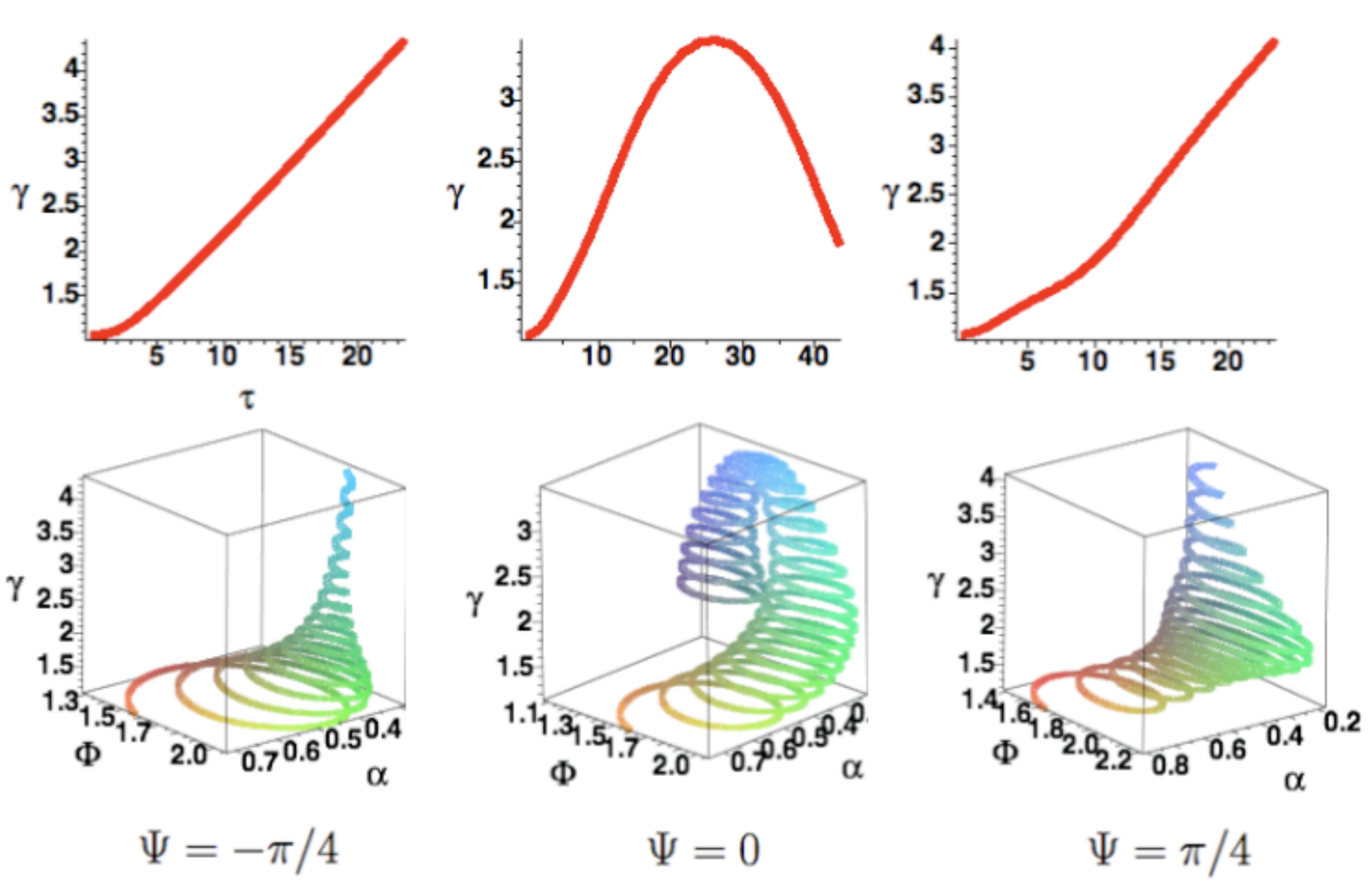}
   \caption{Effect of the phase angle $\psi$ for the longitudinal component on the surfatron process for parameters ($\delta_1=0.06, \delta_2=0.1, n^2=9, \theta=-71^o$). The phase-angle for a given parameter $\eta$ determines whether the longitudinal  parallel component is enabling or breaking the locking of particles into the surfatron.}
    \label{effectpsy}
\end{figure}

\begin{figure}[ht]   \centering
   \includegraphics[width=0.40\textwidth]{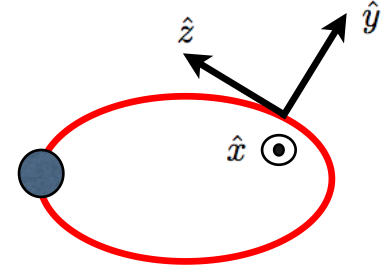}
   \caption{Slab geometry for the dipolar magnetic field}
   \label{topology3}
\end{figure}

\begin{figure}[htb]
\begin{minipage}[b]{0.40\linewidth}
\centering
\includegraphics[width=\textwidth]{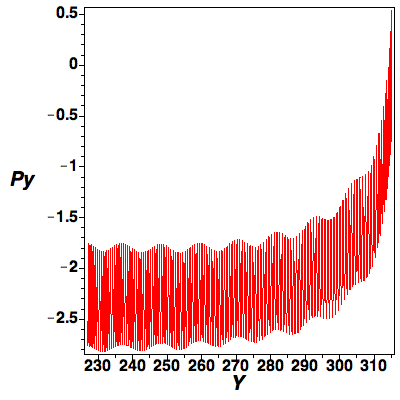}
\end{minipage}
\hspace{0.25cm}
\begin{minipage}[b]{0.40\linewidth}
\centering
\includegraphics[width=\textwidth]{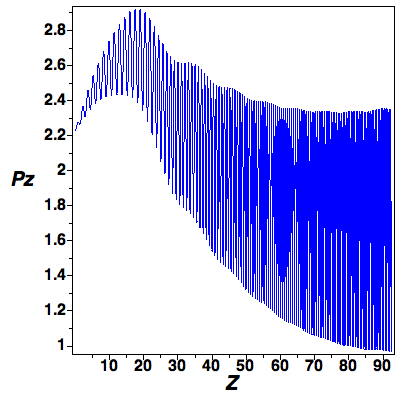}
\end{minipage}
\caption{Orbits for parameters $\delta_1=0.01, v_\phi=0.33 c, \delta_2=0.1$ and  $\theta=71.9^o$. The left panel shows the perpendicular component of momentum $P_y$ against the perpendicular position $Y$. The right panel shows the parallel component of momentum $P_z$ against the parallel position $Z$. Initially the parallel component of momentum increases because of surfatron, until the $-\mu \nabla B$ force becomes sufficiently strong to break the trapping due to the parallel electric field. The particle gains energy of the order of 76 keV during the process.}
\label{gradB_momentum}
\end{figure}

\begin{figure}[ht]
\begin{minipage}[b]{0.40\linewidth}
\centering
\includegraphics[width=\textwidth]{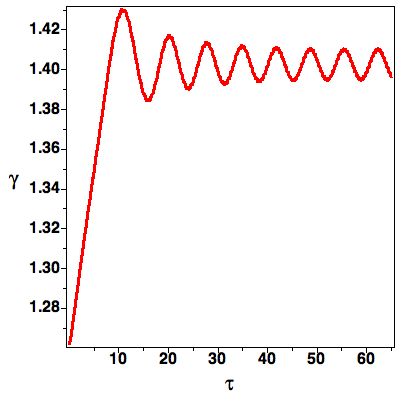}
\end{minipage}
\hspace{0.25cm}
\begin{minipage}[b]{0.40\linewidth}
\centering
\includegraphics[width=\textwidth]{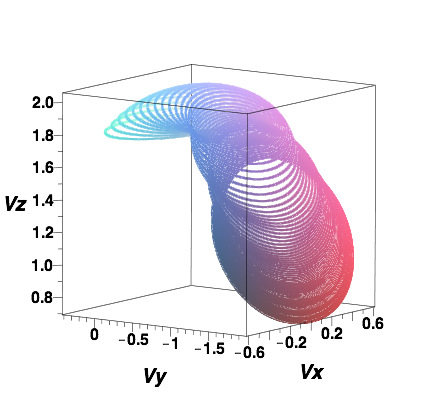}
\end{minipage}
\caption{Orbits for parameters $\delta_1=0.01, v_\phi=0.33 c, \delta_2=0.1$ and  $\theta=71.9^o$. The left panel shows relativistic Lorentz factor $\gamma$ against normalized time $\tau =\omega t$. The right panel shows the three dimensional orbit in velocity space $Vx, Vy, Vz$. For $\tau < 12$, the particle is uniformly accelerated through the surfatron process along the parallel direction. Once the surfatron is made inoperable, the magnetic field gradient dictates the particle orbits and conservation of $\mu$ leads to transfer of energy to the perpendicular direction. The gyroradius of the particle increases ($V_y \gg V_z$) as the particle comes out of the surfatron.}
\label{gradB_gamma}
\end{figure}




\addtocounter{figure}{-1}\renewcommand{\thefigure}{\arabic{figure}a}


\begin{thebibliography}{}
\bibitem[Albert(2002)]{Albert02}
Albert, J. M.: Nonlinear interaction of outer zone electrons with VLF
  waves, Geophys. Res. Lett., 29, 116-1--116-3, \doi{10.1029/2001GL013941}, 2002.

\bibitem[Artemyev et al.(2013)]{Artemyev13}
Artemyev, A.~V., Krasnoselskikh, V.~V., Agapitov, O.~V.,
Mourenas, D. and Rolland, G.: Non-diffusive resonant acceleration of
electrons in the radiation belts, Phys. Plasmas, 19, 122901,
\doi{10.1063/1.4769726}, 2013.

\bibitem[Bell and Inan(1981)]{Bell81}
Bell, T. F. and Inan, U.: Transient nonlinear pitch angle scattering of
energetic electrons by coherent vlf wave packets in the magnetosphere, J.
Geophys. Res., 86, 9047--9063, 1981.

\bibitem[Catell et al.(2008)]{Catell08}
Cattell, C., Wygant, J.~R., Goetz, K., Kersten, K., Kellogg, P.~J., von Rosenvinge,
T., Bale, S.~D., Roth, I., Temerin, M., Hudson, M.~K., Mewaldt, R.~A.,
Wiedenbeck, M., Maksimovic, M., Ergun, R., Acuna, M., and Russell, C.~T.: Discovery of very large amplitude
whistler-mode waves in Earth's radiation belts, Geophys. Res. Lett.
35, L01105, \doi{10.1029/2007GL032009}, 2008.

\bibitem[Chen et al.(2007)]{Chen07}
Chen, Y., Reeves, G. D., and Friedel, R. H. W.: The energization of
relativistic electrons in the outer van allen radiation belt, Nature,
3, 614--617, 2007.

\bibitem[Friedel et al.(2002)]{Friedel02}
Friedel, R. H. W., Reeves, G. D., and Obara, T.: Relativistic electron dynamics
in the inner magnetosphere -- a review, J. Atmos. Sol.-Terr. Phy.,
64, 265--282, 2002.


\bibitem[Green and Kivelson(2004)]{Green04}
Green, J. C. and Kivelson, M. G.: Relativistic electrons in the outer
radiation belt: Differentiating between acceleration mechanisms, J.
Geophys. Res., 109, A03213, \doi{10.1029/2003JA010153}, 2004.

\bibitem[Horne et al.(2005)]{Horne05}
Horne, R. B., Thorne, R. M., Shprits, Y. R., Merredith, N. P., Glauert, S.~A.,
Smith, A. J., Kanekal, S. G., Baker, D. N., Engebretson, M.~J., Posch, J. L.,
Spasojevic, M., Inan, U. S., Pickett, J. S., and Decreau, P. M.: Wave
acceleration of electrons in the van allen radiation belts, Nature, 437, 227--230, 2005.

\bibitem[Kellog et al.(2010)]{Kellog10}
Kellogg, P.~J., Cattell, C.~A., Goetz, K., Monson, S.~J., and Wilson~III, L.~B.: Electron trapping and charge transport
by large amplitude whistlers, Geophys. Res. Lett., 37, L20106, \doi{10.1029/2010GL044845}, 2010.

\bibitem[Kersten et al.(2011)]{Kersten11}
Kersten, K., Cattell, C.~A., Breneman, A., Goetz, K., Kellogg, P.~J., Wygant, J.~R.,
Wilson~III, L.~B., Blake, J.~B., Looper, M.~D., and Roth, I.: Observation of relativistic electron
microbursts in conjunction with intense radiation belt whistler-mode waves,
Geophys. Res. Lett., 38, 8107, \doi{10.1029/2011GL046810}, 2011.

\bibitem[Kulsrud (2005)]{Kulsrud}
Kulsrud, R. M.: Plasma physics for astrophysics, Princeton University
Press, New Jersey, 2005.

\bibitem[Lorentzen et al.(2001)]{Lorentzen01}
Lorentzen, K. R., Blake, J. B., Inan, U. S., and Bortnik, J.: Observations of
relativistic electron microburts in association with vlf chorus, J. Geophys.
Res., 106, 6017--6027, 2001.

\bibitem[Miyoshi et al.(2002)]{Obara02}
Miyoshi, Y., Morioka, A., and Obara, T.: Large enhancementof the outer belt
electrons during magnetic storms, Earth Planets Space, 53, 1163--1170, 2002.

\bibitem[O'brien et al.(2003)]{Obrien03}
O'brien, T. P., Lorentzen, K. R., Mann, I. R., Meredith, N. P., Blake, J. B.,
Fennell, J. F., Looper, M. D., Milling, M. K., and Anderson, R. R.: Energization of
relativistic electrons in the presence of ulf power and mev microbursts:
Evidence for dual vlf and vlf acceleration, J.Geophys. Res., 108, 1329,
doi:10:1029/202JA009784, 2003.

\bibitem[Omura~Hikishima et al.(2010)]{Hikishima10}
Omura Hikishima, M. and Summers, D.: Microburst precipitation of
energetic electrons associated with chorus wave generation, Geophys. Res.
Lett., 37, L07103, \doi{10.1029/2010GL042678}, 2010.

\bibitem[Osmane and Hamza(2012a)]{Osmane12}
Osmane, A. and Hamza, A.~M.: Relativistic acceleration of Landau
resonant particles as a consequence of Hopf bifurcations, Phys. Plasmas, 19, 030702,
\doi{10.1063/1.3692234}, 2012a.

 \bibitem[Osmane and Hamza(2012b)]{Osmane12b}
Osmane, A. and Hamza, A.~M.: Dynamical-systems approach to relativistic nonlinear
wave-particle interaction in collisionless plasmas, Phys. Rev.~E, 85, 056410,
\doi{10.1103/PhysRevE.85.056410}, 2012b.

\bibitem[Shprits et al.(2006)]{Shprits06}
Shprits, Y. Y., Thorne, R. M., Horne, R. B., Glauert, S. A., Cartwright, M.,
Russell, C. T., Baker, D. N., and Kanekal, S. G.: Acceleration mechanism responsible for
the formation of the new radiation belt during the 2003 halloween solar
storm, J. Geophys. Res., 33, L05104, \doi{10.1029/2005GL024256}, 2006.

\bibitem[Summers and Omura(2007)]{Summers07}
Summers, D. and Omura, Y.: Ultra-relativistic acceleration of
electrons in planetary magnetospheres, Geophys. Res. Lett., 34, 24205,
\doi{10.1029/2007GL032226}, 2007.

\bibitem[Thorne et al.(2005)]{Thorne05}
Thorne, R. M., Horne, B., Glauert, S., Meredith, N., Shprits, Y. Y., Ss, D., and
Anderson, R.: The influence of wave-particle interactions on
relativistic electron dynamics during storms, in: vol.~Inner Magnetosphere
Interactions: New Perspectives From Imaging, Geophysical Monograph~159,
edited by: Burch, J., Schulz, M., and Spence, H., American Geophysical Union,
Washington, D.C., 2005.

\bibitem[Wilson~III et al.(2011)]{Wilson11}
Wilson~III, L.~B., Cattell, C.~A., Kellogg, P.~J., Wygant, J.~R., Goetz, K.,
Breneman, A., and Kersten, K.: The properties of large amplitude whistler mode
waves in the magnetosphere: Propagation and relationship with geomagnetic activity,
Geophys. Res. Lett., 38, L17107, \doi{10.1029/2011GL048671}, 2011.


\bibitem[Yoon(2011)]{Yoon11}
Yoon, P. H.: Large-amplitude whistler waves and electron acceleration, Geophys.
Res. Lett., 38, L12105, \doi{10.1029/2011GL047893}, 2011.




\end{thebibliography}
\end{document}